\documentclass[referee]{raa}            

\usepackage{graphicx,times}             
\usepackage{natbib}
\usepackage{amssymb,amsmath}
\usepackage[T1]{fontenc}
\bibpunct{(}{)}{;}{a}{}{,}

\begin{document}

   \title{Photometric Analysis of Three Potential Red Nova Progenitors 
}

   \volnopage{Vol.0 (20xx) No.0, 000--000}      
   \setcounter{page}{1}          

   \author{Surjit S. Wadhwa
      \inst{1}
   \and Ain De Horta
      \inst{1}
   \and Miroslav D. Filipovi\'c
      \inst{1}
    \and N.~F.~H. Tothill
        \inst{1}
    \and Bojan Arbutina
        \inst{2}
    \and Jelena Petrovi\'c
        \inst{3}
    \and Gojko Djura\v sevi\'c
        \inst{3}
   }

   \institute{School of Science, Western Sydney University, Locked Bag 1797, Penrith, NSW 2751, Australia; {\it 19899347@student.westernsydney.edu.au}\\
        \and
             Department of Astronomy, Faculty of Mathematics, University of Belgrade, Studentski trg 16, 11000 Belgrade, Serbia\\
        \and
             Astronomical Observatory, Volgina 7, 11060 Belgrade, Serbia\\
\vs\no
   {\small Received~~20xx month day; accepted~~20xx~~month day}}

\abstract{ We present photometric analysis of three  bright red nova progenitor contact binary systems: ASAS J082151-0612.6, TYC 7281-269-1 and TYC 7275-1968-1. The primary components in all three systems are solar-type low mass stars with radii somewhat larger than their Zero Age Main Sequence counterparts. The secondaries, as in most contact binary systems, have radii and luminosities well above their main sequence counterparts. All three have extreme low mass ratios ranging from 0.075 to 0.097 and two have high degrees of contact, in excess of 75\%. All three have mass ratios and separations below the theoretical values for orbital stability. Chromospheric activity, a hallmark of magnetic activity and magnetic braking, considered important in mediating angular momentum loss, is also explored. All three systems demonstrate the O'Connell effect, and all systems require the introduction of star spots for a better light curve solution. In addition, we show that ASAS J082151-0612.6 and TYC 7281-269-1 have a UV colour excess in the range indicating high chromospheric activity. Another measure of potential significant magnetic activity is X-Ray luminosity; TYC 7275-1968-1, and probably also TYC 7281-269-1, have X-Ray luminosity well above other contact binary systems. We conclude that it is likely that all three are unstable and hence are potential merger candidates.
\keywords{binaries: eclipsing -- stars: mass-loss -- techniques: photometric}
}

   \authorrunning{Wadhwa et al}            
   \titlerunning{Three Red Nova Progenitors}  

   \maketitle

%
%
\section{Introduction}           
\label{sect:intro}

Red Novae and Luminous Red Novae are likely caused by mergers of components of low-mass contact binary systems. Stellar mergers are likely to be quite common with estimates by \citet{2014MNRAS.443.1319K} suggesting a Galactic rate of about once every decade. V1309~Sco remains the only confirmed merger event although only in retrospect \citep{2011A&A...528A.114T}. Pre-merger study of potentially unstable systems is critical to our understanding of the orbital dynamics and internal stellar structure of such systems. Several factors, such as total spin and orbital angular momentum \citep{1995ApJ...444L..41R}, angular momentum loss \citep{2012AcA....62..153S} and the extent of contact between the components \citep{1995ApJ...438..887R}  appear to be among the most critical in influencing orbital stability and potential merger. Recently, interest in identification of  contact binary star merger candidates has seen renewed activity with new projects exploring observational \citep{2021MNRAS.502.2879G} and theoretical \citep{2021MNRAS.501..229W} aspects. Among the most studied instability parameters is the link between the mass ratio of the components and orbital stability \citep{1995ApJ...444L..41R, 2007MNRAS.377.1635A, 2009MNRAS.394..501A}. \citet{2016SASS...35..121V} pointed out that detecting merger candidates using traditional period analysis would be rather impractical requiring vast amounts of data acquired over decades. In addition they noted that most sky survey data is unlikely to be useful given the spread of the observations. The detection of potential red novae candidates thus depends firstly on the identification of probable candidates utilising the link between mass ratio and orbital stability and subsequent focused follow up (potentially over many years) observations.

As part of an ongoing project aimed at identifying potential low mass contact binaries with signs of orbital instability we are in the process of systematically reviewing photometric data from various sky surveys to select potential candidates. As part of our initial review of our algorithm we have identified three potential candidates from the All Sky Automated Survey (ASAS) \citep{2002AcA....52..397P}. We perform detailed photometric analysis of these and show that all three systems show features of orbital instability. 
Magnetic and chromospheric activity indicators show that all three systems are active and likely undergoing angular momentum loss from magnetic braking.


\section{Observations}
\label{sect:Obs}

Photometric observations of three recently identified contact binary systems were performed: ASAS J082151-0612.6 (A0821), TYC 7281-269-1 (T7281) and TYC 7275-1968-1 (T7275). A0821($\alpha_{2000.0} = 08\ 21\ 50.91$, $\delta_{2000.0} = -06\ 12\ 38.5$) an ASAS discovery with a period of 0.309334d and V band amplitude of 0.35mag. The system was observed over 2 nights in March 2020. T7281 ($\alpha_{2000.0} = 14\ 07\ 12.98$, $\delta_{2000.0} = -30\ 24\ 43.7$) is another ASAS discovery with a period of 0.363583d and V band amplitude of 0.23mag. The system was observed over 6 nights in June 2020.

T7275($\alpha_{2000.0} = 13\ 15\ 59.56$, $\delta_{2000.0} = -37\ 00\ 17.7$) A Catalina Survey discovery \citep{2017MNRAS.469.3688D} with a period of 0.382807d and V band amplitude of 0.25mag. The system was recorded by ASAS but not formally recognised as a contact binary. The system was observed over three nights in May and June 2020.

All systems were observed using the 0.6m telescope at the Western Sydney University Penrith Observatory using a SBIG 8300T CCD camera and Johnson $B,V$ and $R$ filters. The total number of observations for each filter along with exposure times are summarised in Table 1. The American Association of Variable Star Observers (AAVSO) VPHOT engine was used for plate solving and differential photometry. The VPHOT engine adjusts the aperture radius to 1.5 times the PSF for each image. TYC 4860-905-1 was used as the comparison star and USNO-B1 0838-0176857 as the check star for A0821, 2MASS 14064960-3023173 was used as the comparison star and 2MASS 14073366-3026255 as the check star for T7281 and TYC 7275-1145-1 was used as the comparison and TYC 7275-1587-1 as the check star for T7275. Given the relative brightness of the contact binary systems we chose comparison stars with similar brightness and colour within the available field. We used the AAVSO Photometric All Sky Survey \citep{2018AAS...23222306H} for the magnitudes of the comparison stars. All observations with estimated error (as provided by VPHOT) exceeding 0.01 magnitude were excluded. As only a single time of minima of the primary eclipse for each system was acquired, refinement of the ephemeris was not undertaken and data folded using the ASAS derived periods. We adopted the deeper eclipse as the primary at phase zero and brightest magnitude determined by means of a parabolic fit of the data around phase 0.25 and phase 0.75.  

Photometry was acquired in $B,V$ and $R$ bands for A0821, $V$ and $R$ bands for T7281, and $V$ band for T7275. All three systems displayed moderate O'Connell effect with the secondary maxima fainter by 0.03 mag ($V$) for A0821 and 0.02 mag ($V$) for T7275, while the primary maxima was fainter by 0.03 mag ($V$) for T7281. The basic catalogue and our light curve findings for the three systems are summarised in Table 1.

\begin{table}

   \centering

   \begin{tabular}{|l|l|l|l|l|l|l|l|}
    \hline
        \hfil Name &\hfil Mag Range &\hfil Obs Num &\hfil Exp Time (s) &\hfil Distance (pc) &\hfil  $E$($B$-$V$)&\hfil  $m_{V1}$ &\hfil  $M_{V1}$\\ 
        A0821 & \hfil 11.85 - 12.15 (V) &\hfil 257  & \hfil 30 &\hfil $342.9\pm12.1$&\hfil0.063&12.12&$4.64\pm0.08$ \\ 
         & \hfil 11.56 - 11.85 (R) &\hfil250   &\hfil 25  &\hfil &\hfil&& \\ 
         & \hfil 12.26 - 12.58 (B) &\hfil 255   &\hfil 50  &\hfil &\hfil&& \\ \hline
          T7281 & \hfil 12.08 - 12.40 (V) &\hfil 160  &\hfil 20 &\hfil $426.4\pm11.0$&\hfil0.059&12.37&$4.40\pm0.06$ \\ 
          & \hfil 11.78 - 12.08 (R) &\hfil 239&\hfil 20 &\hfil &\hfil&& \\ \hline
           T7275 & \hfil 11.47 - 11.71 (V) &\hfil 154  & \hfil 25 &\hfil $326.7\pm60$&\hfil0.059&11.67&$4.28\pm0.44$ \\\hline
    \end{tabular}
    \caption{Basic Catalogue and Light Curve Parameters for the Three Systems. $m_{V1}$ is the Apparent Magnitude of the Primary at Mid-Secondary Eclipse and $M_{V1}$ is the Calculated Absolute Magnitude of the Primary.\label{tbl-1}}
    \end{table}
    
\section{Light Curve Analysis and Determination of Absolute Parameters}
 \label{sec:LCA}
\subsection{Light Curve Analysis}
We used the 2009 version of the Wilson-Devinney (WD) light curve code with Kurucz atmospheres \citep{1971ApJ...166..605W, 1990ApJ...356..613W, 1998ApJ...508..308K, 2021NewA...8601565N} to analyse the photometric data. Complete eclipses are seen in all systems, making them suitable for light curve analysis and determination of the photometric mass ratio \citep{2005Ap&SS.296..221T}. Given the presence of the O’Connell effect in all three systems, both unspotted and spotted solutions were modelled. The absolute magnitude of the primary was determined using the mid secondary eclipse apparent visual magnitude along with the distance and reddening as described in \citet{2021MNRAS.501..229W}. The error in the absolute magnitude was estimated from the reported error in the distance. The effective temperature of the primary was fixed, based on absolute magnitude calibrations from \citet{2013ApJS..208....9P}. 
The common envelope suggests that the components should have similar temperatures, so gravity darkening coefficients were fixed, $g_1 = g_2 = 0.32$ \citep{1967ZA.....65...89L}, as were bolometric albedos, $A_1 = A_2 = 0.5$, and simple reflection treatment was applied \citep{1969PoAst..17..163R}. Logarithmic law \citep{2015IBVS.6134....1N} limb darkening co-coefficients for each wavelength were interpolated from \citet{1993AJ....106.2096V}
 
Due to significant correlation between the geometric elements, it is only possible to estimate the mass ratio of a contact binary system (photometrically) in systems with total eclipses \citep{1993PASP..105.1433R,2005Ap&SS.296..221T}. More recently \citet{2021PASP..133h4202L} has shown the photometric mass ratio error to be in the vicinity of 1\% in the presence of total eclipses. The eclipse amplitude and duration constrain the mass ratio and inclination respectively, overcoming these correlations to some extent, and allowing the mass ratio grid search method 
\citep[or "q-Search", ][]{1982A&A...107..197R} to be used to obtain the mass ratio. The method involves obtaining solutions for various fixed values of $q$, with the inclination and potential (fill-out) acting as free parameters. Each solution is then compared to the observed light curve to find the best solution. Light curve solutions of the three systems (simultaneously where multiple bands were acquired) were modelled at mass ratio intervals of 0.01 over the range 0.05 to 0.9. When a preliminary solution was obtained the search was further refined with analysis at mass ratio intervals of 0.001 on either side of the preliminary solution. The best solution was regarded as the likely mass ratio (without spot). A single spot was introduced by way of trial and error based on the preliminary mass ratio solution and the process repeated again with spot parameters (size, temperature factor and location) also acting as free parameters. The best solution from this set represented the mass ratio with spots. In all cases the spotted solution provided a better fit over the unspotted solution. For the final iteration of the WD code, the mass ratio was also allowed to be a free parameter and the reported standard deviation for each of the variable parameters was adopted as the error for each parameter.

The presence of spots may pose a problem with simultaneous multi-band light curve solutions with inconsistencies between the fitted and observed values.The phenomenon is well recognised, historical examples include AG Vir \citep{1990MNRAS.247..632B}, EH Hya \citep{1991AJ....102..688S} and CQ Cep \citep{1997AN....318..267D} while more recent examples include V781 Tau \citep{2006A&A...452..959K}, MW Pav \citep{2015PASP..127..742A} and  HI Pup \citep{2014NewA...31...56U}. In the case of HI Pup the authors were unable to find a simultaneous spotted solution and could only solve the blue band with the introduction of a spot. The WD code assigns the same spot parameter values for all pass-bands, but this may not be correct for active stars, where there is evidence of significant flux variation at different wavelengths with respect to stellar magnetic activity \citep{2012A&A...539A.140B}. Such variations are likely to introduce fitting errors which cannot be corrected for when attempting to solve multiple pass-bands simultaneously. Solving each band separately may potentially lead to different values for parameters such as the mass ratio thus introducing potential uncertainty. Simultaneous solutions, when possible, probably lead to the correct mass ratio at the expense of robust fitting between the observed and modelled light curves. Such fitting errors are evident in two of our systems where multiple bands were solved simultaneously. Introduction of additional spots or third light did not improve the overall fit. 

\begin{figure}
    \label{fig:LCS}
	\includegraphics[width=1.0\columnwidth]{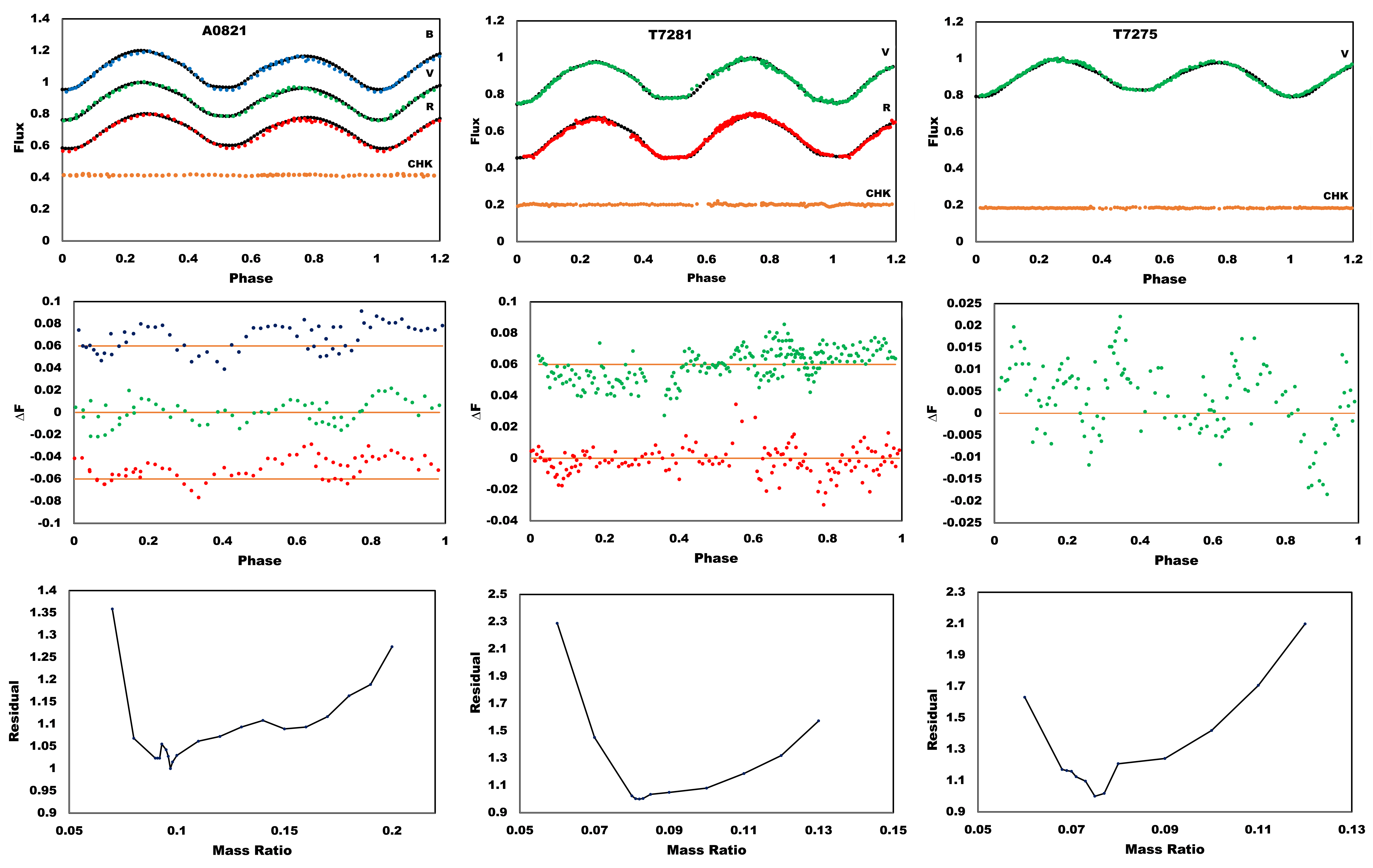}
    \caption{Fitted light curves, residuals and mass ratio search grids for A0821 (left), T7281 (center) and T7275 (right). The upper panel displays the fitted (black) and observed (coloured) light curves of the various bands as labelled. The curves have been normalised to the maximum brightness. In the case of A0821 and T7281 the non V band curves have been shifted vertically for ease of display. Due to the large number of data points the curves for A0821 have been binned 2:1 for the illustration. The middle panel displays the residuals for each band as a difference in the normalised flux. Again where there are multiple bands the curves have been shifted vertically for clarity.  The bottom panel illustrates the mass ratio search grids for the spotted solutions. The residual error has been normalised to the lowest value for clarity. Similarly for clarity the region near the minimum mass ratio is illustrated.}
    \label{fig:P4F1}
\end{figure}

Both T7281 and T7275 were found to be extreme low mass ratio systems ($q = 0.082$ and $q = 0.075$ respectively) with deep contact (>95\% for unspotted solution and >75\% for the spotted solutions). There is good thermal contact between the components with only a small difference in component temperatures. A0821 was also found to be an extreme low mass ratio ($q=0.097$) system in good thermal contact. The degree of contact (38\%) is somewhat lower than the other two systems. The fitted light curves, residuals and mass ratio search grids for the spotted solutions are shown in Figure~\ref{fig:P4F1}, with three dimensional representations of the systems in Figure~\ref{fig:P4F2}. The light curve solutions along with absolute and astrophysical parameters (see below) are summarised in Table 2.

In the case of T7281, the secondary is slightly cooler in the unspotted solution, and slightly hotter in the spotted solution. The primary eclipse however does not change, similar to other low mass ratio systems including \mbox{V1187~Her},  \citep{2017JAVSO..45...11W}, {TY~Pup} \citep{ 2018AJ....156..199S}, {LO~And} \citep{2015IBVS.6134....1N} V857~Her \citep{2005MNRAS.356..765Q}  and \mbox{ASAS~J083241+2332.4} \citep{2016AJ....151...69S}. There was no significant change in the mass ratios for T7281 and A0821 with the addition of a spot; however, the best solution for T7275 resulted in an increase in the mass ratio $q = 0.07$ to $q = 0.075$ with the addition of the spot. Other variations such as change in inclination, temperature of the secondary and degree of contact between the spotted and unspotted solutions are summarised in Table 2. 

\subsection{Absolute Parameters}
 Reliable estimates of the absolute parameters can be made if the mass ratio and the absolute magnitude of the primary are known. The absolute magnitude $M_{V1}$ gives the luminosity of the primary. From the maximum brightness we estimate the luminosity of the system, and hence the luminosity of the secondary, can be estimated. It is reasonably well established that the primary components of contact binary systems can be regarded as  zero-age main sequence (ZAMS) stars \citep{2013MNRAS.430.2029Y}. The mass of the primary was derived from the ZAMS mass-luminosity relation for stars $>$ 0.7$M_{\sun}$ \citep{1991Ap&SS.181..313D}; the mass ratio then yields the mass of the secondary. Errors for the mass of the primary were estimated from the error in the absolute magnitude (see above). The separation of the components $A$ is derived from Kepler's third law, expressed as $(A/R_{\sun})^3=74.5(P/\mathrm{days})^2((M_1+M_2)/M_{\sun})$ with error propagated from the errors in the mass of the components. 

The relative radii of the components are not independent as they are enclosed in a common envelope, but are dictated by the mass ratio and Roche geometry. Using the geometric mean of the fractional radii (from three orientations) provided by the light curve solution we estimate the mean volume fractional radius of the primary and secondary ($r_{1,2}$). From ${R}_1 = r_1 A$ and ${R}_2 = r_2 A$ \citep{2005JKAS...38...43A},  we derive the mean radii of the components for all three systems. As has been reported previously \citep{2012IAUS..282..456S} the relative luminosity and radii of the secondary were considerably higher than those modelled for main sequence counterparts. The absolute parameters based on the best (spotted) light curve solutions are summarised in Table 2.

\begin{figure*}
    \label{fig:3D}
	\includegraphics[width=1.0\columnwidth]{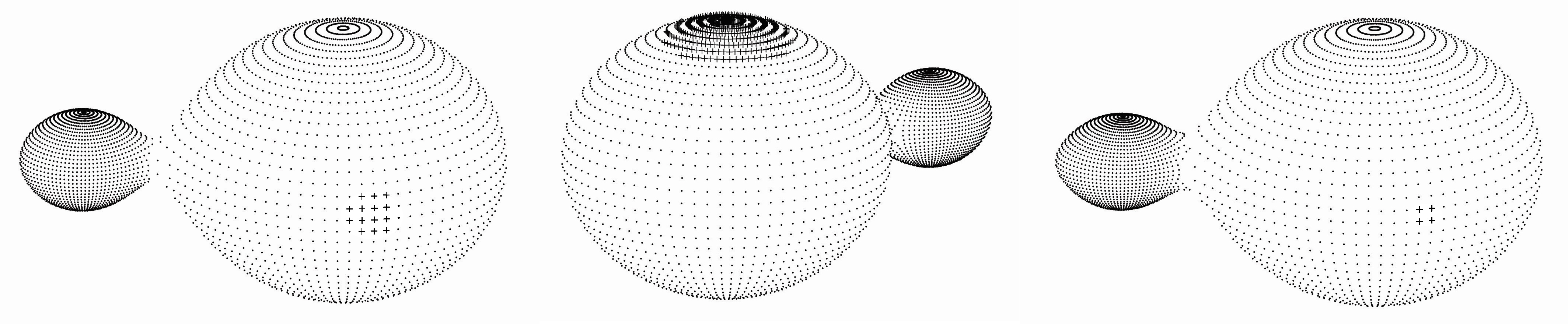}
    \caption{3D representations for ASAS J082151-0612.6 (left), TYC 7281-269-1 (middle) and TYC 7275-1968-1 (right). The spots are shown as the darker regions. The orientation of the 3D images were slightly altered relative to the light curve solutions to allow visualisation of the spots}
    \label{fig:P4F2}
\end{figure*}

\section{Orbital Stability}
\subsection{Instability mass ratio and separation}
Merger of contact binary system is likely when the combined orbital and spin angular momentum of the system is at a critical (minimum) value where tidal instability results as described by \citet{1879Obs.....3...79D}. Recently, \citet{2021MNRAS.501..229W} showed that the mass ratio ($q$) at the onset of orbital instability is dependent on the mass of the primary star for low mass ($0.5M_{\sun}<M<1.6M_{\sun}$) stars. There is no single minimum mass ratio at which instability will occur --- the instability mass ratio is dependent on the gyration radii of the primary and secondary components. The gyration radius for tidally distorted and rotating ZAMS stars of $0.4M_{\sun}<M<1.4M_{\sun}$ is variable and dependent on the mass of the star, so the instability mass ratio is variable and dependent on the mass of the primary.

Briefly, \citet{2021MNRAS.501..229W} showed that by numerically solving  Eq.~(\ref{eq:eq1}) below with appropriate values for the gyration radii, it is possible to determine the mass ratio $q$ at which instability is likely to occur for any given contact binary system:
\begin{equation}
\label{eq:eq1}
\begin{split}
\begin{array}{@{\extracolsep{-3mm}} lll @{}}
\frac{q\frac{k_2^2}{k_1^2}{P}{Q} + \sqrt{(q\frac{k_2^2}{k_1^2}{P}{Q})^2 + 3 (1+q\frac{k_2^2}{k_1^2}{Q}^2) (q \frac{k_2^2}{k_1^2}{P}^2 + \frac{q}{(1+q)k_1^2})   }} {q \frac{k_2^2}{k_1^2}{P}^2 + \frac{q}{(1+q)k_1^2}} 
  -  \frac{0.6q^{-2/3} + \ln (1+ q^{-1/3})}{0.49q^{-2/3}+0.15 f}  = 0,
 \end{array}
 \end{split}
\end{equation}
where $k_{1,2}$ are the gyration radii of the primary and secondary, $f$ is the fill-out factor ($0<f<1.0$) and $P$ and $Q$ are defined similarly to \citet{2007MNRAS.377.1635A}:
\begin{footnotesize}

\begin{equation}
P = \frac{0.49q^{2/3}-3.26667q^{-2/3}(0.27q -0.12q^{4/3})}{0.6q^{2/3} + \ln (1+ q^{1/3})} ,
\end{equation}
\end{footnotesize}
and
\begin{footnotesize}
\begin{equation}
Q =\frac{(0.27q -0.12q^{4/3})({0.6q^{-2/3} + \ln (1+ q^{-1/3})})}{0.15 (0.6q^{2/3} + \ln (1+ q^{1/3}))}.
\end{equation}
\end{footnotesize}

If the Roche distorted radius of the primary is known, it is possible to determine the separation at which instability is likely \citep{2007MNRAS.377.1635A} by:
\begin{small}

\begin{equation}
\label{eq:a-inst}
 \frac{A_{\mathrm{\scriptscriptstyle inst}}}{R_1} = \frac{q\frac{k_2^2}{k_1^2}{P}{Q} + \sqrt{(q\frac{k_2^2}{k_1^2}{P}{Q})^2 + 3 (1+q\frac{k_2^2}{k_1^2}{Q}^2) (q \frac{k_2^2}{k_1^2}{P}^2 + \frac{q}{(1+q)k_1^2})   }}{q \frac{k_2^2}{k_1^2}{P}^2 + \frac{q}{(1+q)k_1^2}}.
\end{equation}
\end{small}

\citet{2009A&A...494..209L} modelled values of the gyration radii $k$ of low mass ($0.09M_{\sun} < M < 3.8M_{\sun}$) rotating and tidally distorted zero age main sequence (ZAMS) stars. We use the data to derive a linear fit for $0.6_{\sun} < M < 1.4M_{\sun}$ as follows:

\begin{equation}
    k_1=-0.2392M_1 + 0.527,
\end{equation}
that we use to calculate gyration radius of the primary.

The mass ratio of contact binaries approaching merger is low and so the mass of the secondary is very small. For very low mass stars the totally convective $n=1.5$ polytrope should be appropriate, with $k^2\approx 0.205$ \citep{2007MNRAS.377.1635A}. Nevertheless, we use \citet{2009A&A...494..209L} data for $0.09_{\sun} < M < 0.2M_{\sun}$ stars and derived a linear fit in this range
\begin{equation}
    k_2=-0.1985M_2 + 0.485,
\end{equation}
which we use for the secondary. We combine these gyration radii  to solve  Eqs.~(\ref{eq:eq1}) and (\ref{eq:a-inst}), yielding the instability mass ratio $q_{inst}$ and instability separation $A_{inst}$ for each system. We used $q_{inst}$ rather than the light curve solution mass ratio to calculate the instability separation. All three systems have mass ratios and separations below the critical values where orbital instability is likely, so they are potential merger candidates. The instability parameters are summarised in Table 2.

\begin{table*}
\footnotesize
    \centering
           \begin{tabular}{|l|l|l|l|l|l|l|}
    \hline
           & A0821   &  & T7281 &  & T7275 &  \\
         Parameter & No Spot &Spot & No Spot &Spot &No Spot &Spot \\ \hline
         $T_1$ (K) (Fixed) & 5850 &5850 &5960 &5960& 6010 &6010 \\
         $T_2$ (K) & $5710\pm40$ & $5705\pm44$ & $5840\pm45$ & $6240\pm38$ & $5765\pm86$ &$5615\pm65$ \\
        Inclination ($^\circ$) & $70.06\pm1.50$ &$70.1\pm1.4$ &$74.4\pm2.0$ &$74.2\pm1.0$ &$68.8\pm1.4$ &$69.8\pm1.5$ \\
        Fill-out (\%) & $38\pm4$ &$35\pm4$ &$99\pm1$ &$77\pm2$ &$99\pm1$ &$83\pm2$ \\
        $r_1$ (mean) &  &0.6288 &  & 0.6170 &  & 0.5242 \\
        $r_2$ (mean) &  &0.2458 &  & 0.2230 &  & 0.2195 \\
        $q$ ($M_2/M_1$) & $0.097\pm0.010$ & $0.097\pm0.010$ & $0.082\pm0.004$ & $0.082\pm0.004$ & $0.070\pm0.002$ &$0.075\pm0.002$ \\
        $q_{inst}$ & 0.107 & 0.107 &0.105 & 0.101 & 0.099 & 0.098 \\
        $A$/$R_{\sun}$ & $1.99\pm0.01$ & $1.99\pm0.01$ & $2.25\pm0.01$ & $2.25\pm0.01$ & $2.35\pm0.01$ & $2.35\pm0.01$ \\
        $A_{inst}$/$R_{\sun}$ & -- & 2.00 &  -- & 2.29  &  --  & 2.41 \\
        $M_1/M_{\sun}$ &$1.01\pm0.01$  & $1.01\pm0.01$ & $1.08\pm0.01$ & $1.08\pm0.01$ & $1.11\pm0.02$ & $1.11\pm0.02$ \\
        $M_2/M_{\sun}$ &$0.098\pm0.002$  & $0.098\pm0.00$2 & $0.088\pm0.002$ & $0.088\pm0.002$ & $0.078\pm0.002$ & $0.084\pm 0.002$ \\
        $R_1/R_{\sun}$ &  & $1.18\pm0.02$ &  & $1.39\pm0.02$ &  & $1.47\pm0.02$ \\
        $R_2/R_{\sun}$ &  & $0.43\pm0.01$ &  & $0.50\pm0.02$ &  & $0.52\pm0.021$ \\
        $M_{V1}$ &  & $4.64\pm0.08$ &  &$4.40\pm0.06$ &  &$4.28\pm0.44$ \\
        $M_{FUV}$ &  &14.66 &  &12.82 &  &  \\
        $M_{FMS}$ &  &16.84 &  &16.01 &  &  \\
        UV Colour Excess  &  &-1.71 &  &-2.69 &  & \\ \hline
        Spot Details&&&&&&\\
        Latitude& &90&&120&&90\\
        Longitude& &260&&10&&250\\
        Radius ($^{\circ}$)& &$10.2\pm0.6$&&$26.7\pm0.77$&&$5.01\pm0.17$\\
        Temp Factor& &$1.2\pm0.05$&&$1.10\pm0.01$&&$1.38\pm0.03$\\ \hline
    \end{tabular}
    \caption{Light curve solution, absolute parameters and instability parameters for ASAS J082151-0612.6, TYC 7281-269-1 and TYC 7275-1968-1. Radii , $A_{inst}$, absolute magnitudes and color access are only shown for the best fitting (spotted) solutions}
\end{table*}

\subsection{Magnetic Activity}
Loss of angular momentum from binary systems causes the separation between the components to decrease \citep{2004MNRAS.355.1383L}, and the most likely mechanism for angular momentum loss in contact binary systems is magnetic braking. There is indirect evidence that angular momentum loss from cool contact binaries occurs by way of magnetic stellar winds. Increased chromospheric activity and/or strong ultraviolet emissions are a hallmark of increased magnetic activity and thus increased potential for magnetic braking and magnetic stellar winds in contact binary systems \citep{2004MNRAS.355.1383L, 1983MNRAS.202.1221R, 1983HiA.....6..643V}. There are many magnetic and chromospheric signals associated wih contact binaries. The asymmetry in the maxima (O’Connell effect) is the most easily observed light curve feature. Light curve analysis, apart from incorporating star spots, provides little indication of chromospheric activity. Certain spectral emission lines particularly in the high energy bands provide much clearer indicators of enhanced chromospheric activity. Chromospheric emissions in the visual band are swamped by the light of the photosphere; but this is not the case in the far-ultraviolet region, especially in the case of low mass dwarfs common amongst contact binary systems \citep{2010PASP..122.1303S}. The GALEX (Galaxy Evolution Explorer) satellite imaged the sky in both the far-ultraviolet band (FUV) centered on 1539{\AA} and a near-ultraviolet band (NUV) centered on 2316{\AA}. Unfortunately the NUV band may be contaminated by photospheric flux and only the FUV band can be relied upon for the detection of chromospheric activity \citep{2010PASP..122.1303S}. 

 The fraction of a stars emission in the active H and K lines \citep{1984ApJ...279..763N} is an excepted measure of chromospheric emission strength. This fraction is normally expressed as $\log R_{HK}^{\prime}$,
 with $\log R_{HK}^{\prime} \geq -4.75$ characteristic of a more active star  \citep{1996AJ....111..439H}.
 Using the large database of $\log R_{HK}^{\prime}$ values of dwarf stars produced by the Mount Wilson HK project \citep{1978PASP...90..267V, 1991ApJS...76..383D},
 \citet{2010PASP..122.1303S} matched GALEX FUV magnitudes ($m_{FUV}$) to the $\log R_{HK}^{\prime}$ for dwarf stars to derive two key relationships:
\begin{equation}
    (m_{FUV}-B)_{base} = 6.73(B-V) + 7.43,
\end{equation}
which defines $m_{FUV}-B$ for stars with the weakest emissions and low activity;
and
\begin{equation}
    \Delta(m_{FUV}-B) = (m_{FUV}-B) - (m_{FUV}-B)_{base}
\end{equation}
defining a colour excess.
For active stars, i.e.~those with $\log R_{HK}^{\prime} \geq -4.75$, they deduced that the colour excess was essentially always less than $-0.5$ and often less than $-1.0$. The opposite was true for less active stars with colour excesses significantly higher than $-0.5$. In a somewhat broader study, \citet{2011AJ....142...23F} detected a good correlation between high $\log R_{HK}^{\prime}$ values and higher FUV luminosity, and in an earlier study \citet{2010AJ....139.1338F} estimate  absolute FUV magnitudes ($M_{FMS}$) for single main sequence stars from B8 to M2 spectral types. 

All three systems show the O'Connell effect, suggesting the possibility of star spots and all three have a better light curve fit when at least one spot is introduced. Two of the systems were observed by GALEX, so a more detailed examination of the chromospheric/magnetic activity is possible. Reported apparent FUV magnitudes for each system were converted to absolute magnitudes $M_{FUV}$ and compared with the absolute GALEX magnitude for single stars ($M_{FMS}$) of same mass. $M_{FUV}$ for A0821 was 2.18 magnitudes brighter than the corresponding main sequence star, and for T7281 absolute FUV magnitude was 3.19 magnitudes brighter. The UV excess (using published values for $B$ magnitude) for A0821 is estimated to be $-1.71$ and for T7281, $-2.69$. Both significantly less than -1.0 considered the hallmark for chromospherically active stars. The chromospheric activity parameters are summarised in Table 2.

W UMa stars are known X-ray emitters. The X-ray emission mechanism is not well understood but may relate to the synchronous fast rotation of the components and magnetic activity related to the common convective envelope \citep{2004A&A...426.1035G}. \citet{2001A&A...370..157S} reported on the X-ray properties of 57 contact binary systems from the ROSAT survey and found a  median X-ray luminosity ($\log L_X$) of 30.04, with mean of 30.18. Only one of our systems was detected as a point source by ROSAT. T7275 is less than 20 arcsec from a ROSAT faint point source \citep{2000IAUC.7432....3V} with a count rate $3.18\times10^{-2}$\,s$^{-1}$, which equates to a flux $f_X= 3.36\ \times 10^{-13}$\,erg\,cm$^{-2}$cts$^{-1}$ after applying the energy conversion factor detailed in \citet{2001A&A...370..157S}. Using
\begin{equation}
    L_X = 4{\pi}d^2f_X
\end{equation}
where $d$ is the distance, we determined $\log L_X = 30.63$ for T7275, somewhat higher than the mean reported by \citet{2001A&A...370..157S}, suggesting a more active system.

Two sources in the Swift point source catalog \citep{2020ApJS..247...54E} coincide to within 2 arc seconds to the positions of T7275 and T7281 and are likely associated with the contact binary systems. X-ray luminosities for each system using the the mean 0.3--10\,keV observed flux (power law spectrum) are 30.23 and 30.52, respectively. Although not directly comparable to the ROSAT flux, the X-ray luminosities appear to be significantly elevated above average, again suggesting both systems to be magnetically active and subject to enhanced magnetic braking and angular momentum loss.

\section{Summary and Conclusions}
\label{sec:summary}
We present initial photometric analysis of three recently identified bright contact binary systems. All three have extremely low mass ratios ranging from 0.075 to 0.097 with two having very high degree of contact. Estimates of the absolute parameters of the components were made possible by well-established parallax distances. The primary in all cases exhibits properties similar to other ZAMS stars of similar mass, while the secondaries, as is typical in contact binary systems, have larger radius and luminosity than their main sequence counterparts. 

Dynamic stability of contact binaries is dependant upon the interplay between spin angular momentum and the orbital angular momentum. Stability is only possible if the orbital angular momentum is roughly more than three times the total spin angular momentum \citep{1980A&A....92..167H}. As it was shown  \citep{1995ApJ...444L..41R, 2006MNRAS.369.2001L, 2007MNRAS.377.1635A, 2009MNRAS.394..501A, 2021MNRAS.501..229W} there is a direct link between the mass ratio of a contact binary system and the spin and orbital angular momentum. Orbital instability cannot occur if the mass ratio is above a critical value. The critical mass ratio is dependant on the mass of the primary and unique for each  system \citep{2021MNRAS.501..229W}. In order to determine if any of the systems is a potential merger (red nova) candidate, we derive the theoretical instability mass ratio and separation for each system, compare them with current mass ratios and separations derived from the light curve analysis, and show that all three exhibit criteria suggestive of orbital instability.

As noted by \citet{2004MNRAS.355.1383L} the most likely mechanism of angular momentum loss (and hence potential orbital instability) in contact binary systems is magnetic braking. Ultraviolet and X-Ray observations of contact binary systems \citep{1983MNRAS.202.1221R, 1983HiA.....6..643V} have shown that contact binary systems are strong  UV and X-Ray sources indicating the presence of a magnetic field and probable active magnetic breaking. We look at the chromospheric and magnetic activity as a defacto marker for potential angular momentum loss and find that all three systems have at least some features of enhanced chromospheric activity. Two systems observed by GALEX show signs of significant chromospheric activity with absolute FUV magnitudes significantly brighter than comparable main sequence single stars and UV colour excess well into the range considered to represent increased activity. Two systems also show evidence for significant X-ray emission, also a marker of significant chromospheric and magnetic activity.

Variation in period has been suggested as a marker of orbital instability, however, variation is commonly observed, and as pointed out by \citet{2018MNRAS.474.5199L}, both increasing and decreasing variations in contact binary periods are observed equally. The variations maybe be due to apsidal motion, light travel time effects, magnetic activity cycles, mass transfer/loss, while the long term period decrease is usually due to angular momentum loss \citep{2018MNRAS.474.5199L}. None of the systems reported in this study have historical high cadence observations to mount a meaningful short term period study analysis as a marker for orbital instability. In addition, due to the competing mechanisms just mentioned, using period study as an identifier of potential orbital instability is quite impractical.

Given these findings, we believe that all three systems are likely to be unstable and are potential red nova progenitors. The time frame from the onset of instability to eventual merger is unknown. Photometric data for V1309 Sco were available for approximately 6 years prior to the merger event, and among the most critical changes determined retrospectively were the exponential decline in the period of the V1309 Sco progenitor particularly in the three years prior to merger and the shape of the light curve, especially with respect to the secondary maxima in the six years prior to the merger event \citep{2011A&A...528A.114T}. Identification of potential merger candidates and their study prior to the merger event is critical to our understanding of contact binary evolution. Given the relative brightness of these three potential red nova progenitors, long term follow-up monitoring of period and light curve changes can be made with modest equipment and could potentially be performed by advanced amateurs through organisations such as the AAVSO.

\section*{Acknowledgements}
Based on data acquired on the Western Sydney University, Penrith Observatory Telescope. We acknowledge the traditional custodians of the land on which the Observatory stands, the Dharug people, and pay our respects to elders past and present.

B.A. acknowledges the financial support of the Ministry of
Education, Science and Technological Development of the Republic
of Serbia through contract No. 451-03-68/2022-14/200104.

J. P. and G. D.  gratefully  acknowledge financial support of the Ministry of Education, Science and Technological Development of the Republic of Serbia through contract No. 451-03-9/2021-14/200002.

This research has made use of the SIMBAD database, operated at CDS, Strasbourg, France.
\bibliographystyle{raa}
\bibliography{bibtex}

\end{document}